\documentclass[superscriptaddress,showpacs,aps,twocolumn,amsmath,amssymb,prl]{revtex4}
\usepackage{graphicx}
\usepackage{dcolumn}
\usepackage{bm}
\usepackage{textcomp}
\usepackage{verbatim} 

\usepackage{color}
\definecolor{darkblue}{rgb}{0,0,0.5}

\begin{document}


\title{Nonlinear Acoustics at GHz Frequencies in a Viscoelastic Fragile Glass Former}

\author{Christoph Klieber}
\email{klieber@mit.edu}
\altaffiliation[Current address: ]{EP Schlumberger, 1 rue Henri Becquerel, 92140 Clamart, France}
\affiliation{Department of Chemistry, Massachusetts Institute of Technology, Cambridge, MA 02139, USA}
\affiliation{Institut Mol\'ecules et Mat\'eriaux du Mans, UMR-CNRS 6283, Universit\'e du Maine, 72085 Le Mans, France}
\author{Vitalyi E. Gusev}
\affiliation{Institut Mol\'ecules et Mat\'eriaux du Mans, UMR-CNRS 6283, Universit\'e du Maine, 72085 Le Mans, France}
\author{Thomas Pezeril}
\affiliation{Institut Mol\'ecules et Mat\'eriaux du Mans, UMR-CNRS 6283, Universit\'e du Maine, 72085 Le Mans, France}
\author{Keith A. Nelson}
\email{kanelson@mit.edu}
\affiliation{Department of Chemistry, Massachusetts Institute of Technology, Cambridge, MA 02139, USA}

\date{\today}

\begin{abstract}
\noindent Using a picosecond pump-probe ultrasonic technique, we study the propagation of high-amplitude, laser-generated longitudinal coherent acoustic pulses in the viscoelastic fragile glass former DC704. We observe an increase of almost ten percent in acoustic pulse propagation speed of its leading shock front at the highest optical pump fluence which is a result of the supersonic nature of nonlinear propagation in the viscous medium. From our measurement we deduce the nonlinear acoustic parameter of the glass former in the GHz frequency range across the glass transition temperature.
\end{abstract}


%

\maketitle

Observation of laser-driven shock wave propagation provides direct experimental access to the equation of state of strongly compressed materials. This information is of paramount importance for geophysics, astrophysics \cite{SD08} and inertial confinement fusion \cite{CSC+98}. For many years, measurements of shock velocities have been possible through transit time measurements \cite{KDD08}. Only recently, single-shot optical velocity interferometry \cite{CSC+98, CCS+98, BKB+99, BEH+04, RGB+07, RMC+10, BGS+11} allowed direct access to the dynamics of shock front motion in transparent solids but has been restricted to experimental configurations where the shock transforms the media into a new, highly reflecting phase. Through this technique, the decay of both plane \cite{CCS+98, RGB+07} and convergent \cite{BGS+11, PSV+11} shocks with pressures exceeding tens of GPa have been reported. The propagation of shock waves in soft transparent materials such as polycarbonate and PMMA has been observed by single-shot ultrafast dynamic ellipsometry \cite{MMF03, BMMF07}, a method allowing the separation of pressure-induced variations in elastic and optical properties. In such materials, characteristic pressures ranged from a few GPa to 10-20~GPa and acoustic Mach numbers defined as $M_A=u/v_0$ -- where $u$ is the particle velocity and $v_0$ the linear acoustic velocity -- were subsonic in the range $0.2<M_A<0.9$. Very recently, the propagation of weak shock waves with Mach numbers $M_A<0.1$ and pressures below the damage threshold of the sample has been observed in thin sapphire slabs through ultrafast optical reflectivity \cite{VCD06}, in a 4:1 methanol-ethanol mixture in a diamond anvil cell by ultrafast velocity interferometry \cite{ACRZ08}, in a piezoelectric thin film through THz spectroscopy \cite{armstrong09}, in a gold film through ultrafast plasmon interferometry \cite{temnov13} and ultrafast optical imaging \cite{pezeril14}. In such weakly nonlinear acoustics experiments \cite{RS77}, no variation of the weak shock wave velocity during propagation has been reported to date. In a different manner, an indication for the nonlinearity of picosecond acoustic pulses with Mach numbers $0.0007<M_A<0.0018$ and their decay within 1-3~mm propagation distances has been observed by classical Brillouin spectroscopy \cite{MD02} through measurement of the distribution of 22~GHz longitudinal phonons along the acoustic pulse trajectory. In such experiments, information on wide-frequency-band nonlinear waves is obtained from the spatial distribution of its single Brillouin frequency component. The dependence of the Brillouin frequency shift on the amplitude/velocity of nonlinear acoustic pulses has not been reported so far.

In this letter we report direct measurements of the velocity of decaying, weak shock fronts using the technique of picosecond Time-Domain Brillouin Scattering (TDBS) \cite{TGMT86}. To our knowledge, this technique has so far only been applied to measurements of variations in the propagation velocities of linear acoustic pulses caused by the spatial inhomogeneous heating of a material or elastic inhomogeneities of the material itself along the propagation direction \cite{MRB+09}. In this work we directly observe the decay of a weak shock wave as it propagates. Variations of the shock velocity with propagation distance and across the glass transition temperature are revealed through variations in the Brillouin scattering frequency. Our observations lead to the estimate of the nonlinear acoustic parameter of a fragile glass former \cite{Ang95} DC704 at around 20~GHz at temperatures across the glass transition.

\begin{figure}[t!]
\centering
\includegraphics[width=8.5cm]{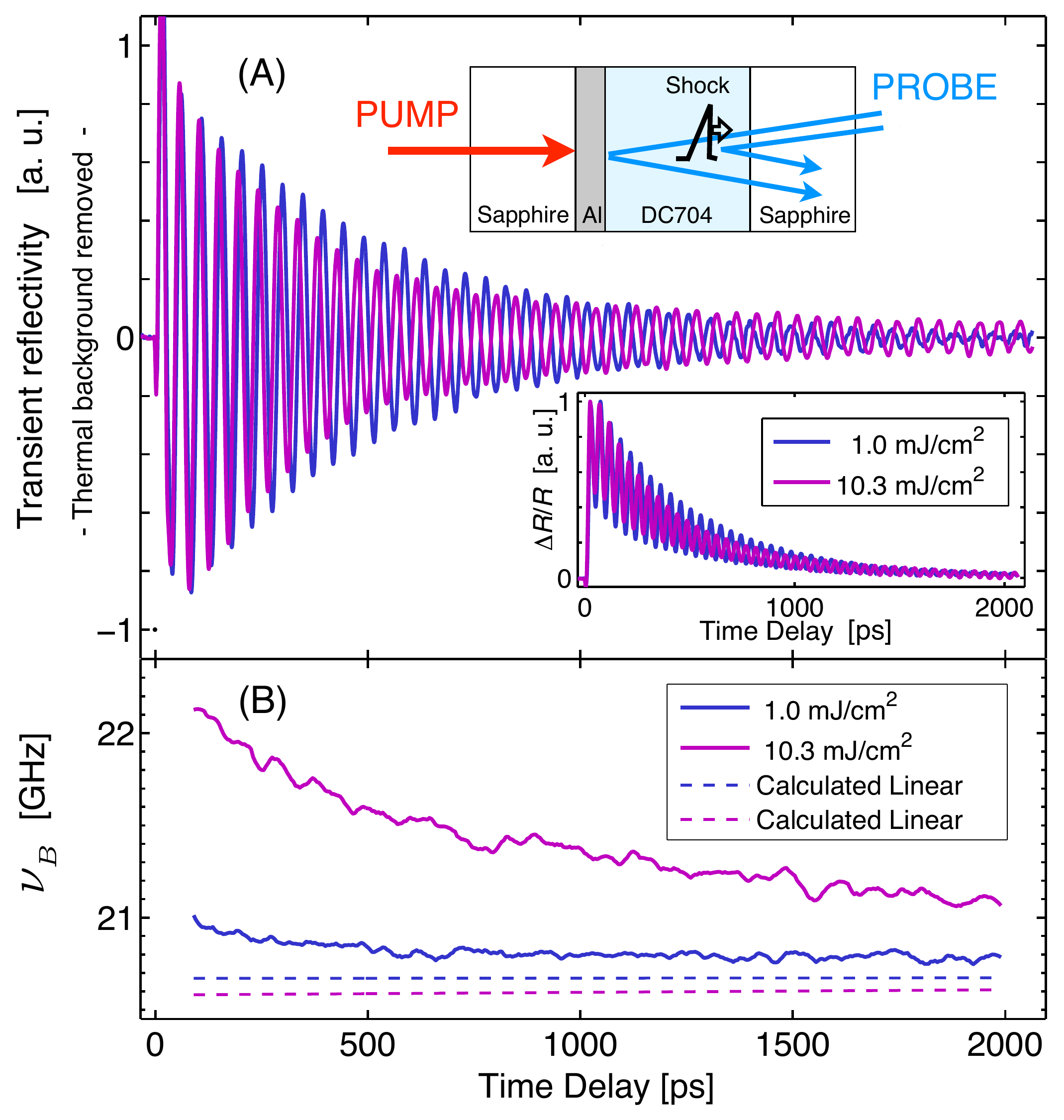}
\caption{\label{fig:BS_rawdata} (Color online) (A) Inset (top): Experimental setup with the $\sim$100~$\mu$m liquid squeezed in between two sapphire substrates, one of them holding a 33~nm aluminum photoacoustic transducer film that launched an acoustic wavepacket into the liquid through transient absorption of an optical pump pulse. Acoustic propagation through the liquid was then detected by a time-delayed optical probe pulse. (A) Inset (bottom): The propagating acoustic wavepacket in the transparent liquid resulted in Time-Domain Brillouin Scattering oscillations. (A) Normalized datas recorded at 200 K at two representative low and high fluences with thermoreflectance background removed. (B) Extracted values for the time variation of the \textquotedblleft local\textquotedblright~Brillouin frequency $\nu_{_B}$. Clearly apparent is the frequency down-chirp with increasing time at high fluence caused by nonlinear acoustic pulse propagation of the weak shock.}
\end{figure}

Samples were prepared by squeezing liquid tetramethyl tetraphenyl trisiloxane (trade name DC704, glass transition temperature \cite{JNO05} $T_g\approx$~210~K, commonly used as diffusion pump oil) between two optically clear substrates, a generation side substrate which held a 33~nm aluminum transducer film and a detection side substrate, see inset Fig.~\ref{fig:BS_rawdata}(A). The liquid thickness was about 100~$\mu$m. Anhydrous DC704 was used as purchased from Sigma-Aldrich and forced through several linked 0.2~$\mu$m teflon millipore filters to remove dust particles before applying to the sample without further purification. After the sample was built it was transferred to a cryostat and the sample chamber was immediately evacuated.

Our front-back pump-probe setup was based on the common picosecond ultrasonics approach based on TDBS \cite{Mar98,MTZM96,pezeril09,pezeril12}. Absorption of an optical pump pulse and subsequent rapid thermal expansion launched the longitudinal acoustic wavepacket into the sample. The excitation pulses from a Ti:Sapphire amplifier laser system (Coherent RegA, 250~kHz repetition rate, 790~nm wavelength, 8~nm bandwidth, 200~fs pulse duration), were focused on the sample to a 100~$\mu$m spot. A small portion of the laser output was frequency doubled to 395~nm wavelength and time-delayed to serve as a probe. Expansion of the laser beam to about 1~cm diameter allowed us to ensure good beam pointing stability and practically no change of the beam diameter of 40~$\mu$m at the sample.

Propagation of the acoustic waves from the transducer film into the liquid were optically detected by TDBS. The coherently scattered field, whose optical phase varied depending on the acoustic wave peak and null positions, superposed with the reflected probe field, resulting in signal intensity that showed time-dependent oscillations. The frequency $\nu_{_B}$ of these oscillations is related as in any Brillouin scattering measurement to the propagation velocity $v$ of the Fourier component $\nu_{_B}$ of the acoustic field, to the index of refraction $n$ at the probe wavelength $\lambda$, through the relation (in case of normal incidence of the probe beam),
\begin{eqnarray}
\label{nu}
\nu_{_B} = 2 n v / \lambda \, .
\end{eqnarray}
Coherent Brillouin scattering data obtained at two different pump laser fluences are shown in Fig.~\ref{fig:BS_rawdata}(A). A nonoscillatory signal component due to thermoreflectance of the aluminium film has been subtracted to emphasize the acoustic signal components. The routine fitting procedure based on two damped exponents that we used to filter the thermal background did not influence the Brillouin component. After background removal, we fitted the whole time interval with an exponentially damped sinusoidal form to obtain a precise value of the initial input Brillouin frequency. In a second step we selected a short time interval, about two and one half oscillation cycles, and fitted it with adjustable values for the Brillouin frequency $\nu_B$. We associated the obtained frequency with the middle value of the time interval. After shifting the time interval by 10~ps to larger times, we repeated the second step over the whole available time delay between about 25~ps and 2000~ps, which allowed to extract the time variation of the \textquotedblleft local\textquotedblright~Brillouin frequency. As a check on our procedure, we plotted the variable-frequency function determined by the successive fits and compared it to the data set. The fitted functions coincide almost exactly with the data set. Fig.~\ref{fig:BS_rawdata}(B) shows the fitting results, the time evolution of the \textquotedblleft local\textquotedblright~Brillouin frequency. At the relatively low pump fluence of 1.0~mJ/cm$^2$, the acoustic strain amplitude is close to the limit of the linear response and the Brillouin frequency is therefore almost constant over the recorded time delay interval. On the other hand, at about ten times higher pump fluence of 10.3~mJ/cm$^2$, the initial Brillouin scattering frequency is significantly higher than in the low pump fluence limit and decreases with increasing time delay. This Brillouin frequency down-chirp is a result of nonlinear acoustic effects. However, since the temperature rise can induce a change in the refractive index which may affect the Brillouin scattering frequency (from Eq.~(\ref{nu})), careful simulations of heating effects \cite{Kli10} in the sample, both on a single shot basis (fs to ps time scales) and steady state, were carried out in order to avoid erroneous interpretation of the results. Our simulations of the heat diffusion reveal that the efficient heat flow into the sapphire allowed rapid cooling of the temperature rise at the laser excited aluminum-sapphire interface to just 10~\% of its initial value (which could be as large as several hundred Kelvin) after $\sim$30~ps. Thus, single-shot heat flow into the liquid was negligible, and had no detectable effect on the signal. In addition, subsonic heat flow into the liquid has no influence on the propagating acoustic pulse and therefore, single-shot heating effects on the sound speed were irrelevant. In contrast, cumulative heating caused a slight homogenous temperature rise in the liquid of up to 4.5~K at the highest pump fluence of 10.3~mJ/cm$^2$, which is relatively low and can't explain the observed change in Brillouin scattering frequency.

In fact due to nonlinear multi-frequency interaction of all the spectral components of the large-frequency-band weak shock pulse, the high frequencies components, including the specific Brillouin frequency $\nu_{_B}$, become preferentially spatially localized at the vicinity of the weak shock front. This is known as the process of nonlinear steepening of the shock front during propagation. Consequently, the velocity related to the Brillouin frequency component matches the velocity of the weak shock front. This qualitatively explains the sensitivity of this measurement to the velocity of the weak shock front through TDBS, which in turn depends on the amplitude of the shock. Since the light scattering process is predominant at the moving shock front, which can be approximated as a moving weakly reflecting mirror, the Brillouin frequency $\nu_{_B}$ corresponds to the Doppler frequency shift of the reflected probe light, proportional to the velocity of the mirror. Then Brillouin scattering results reveal the change in the weak shock speed $v$ of equation Eq. (\ref{nu}) induced by nonlinear acoustic effects at the frequency $\nu_{_B}$.

\begin{figure}[t!]
\centering
\includegraphics[width=8.5cm]{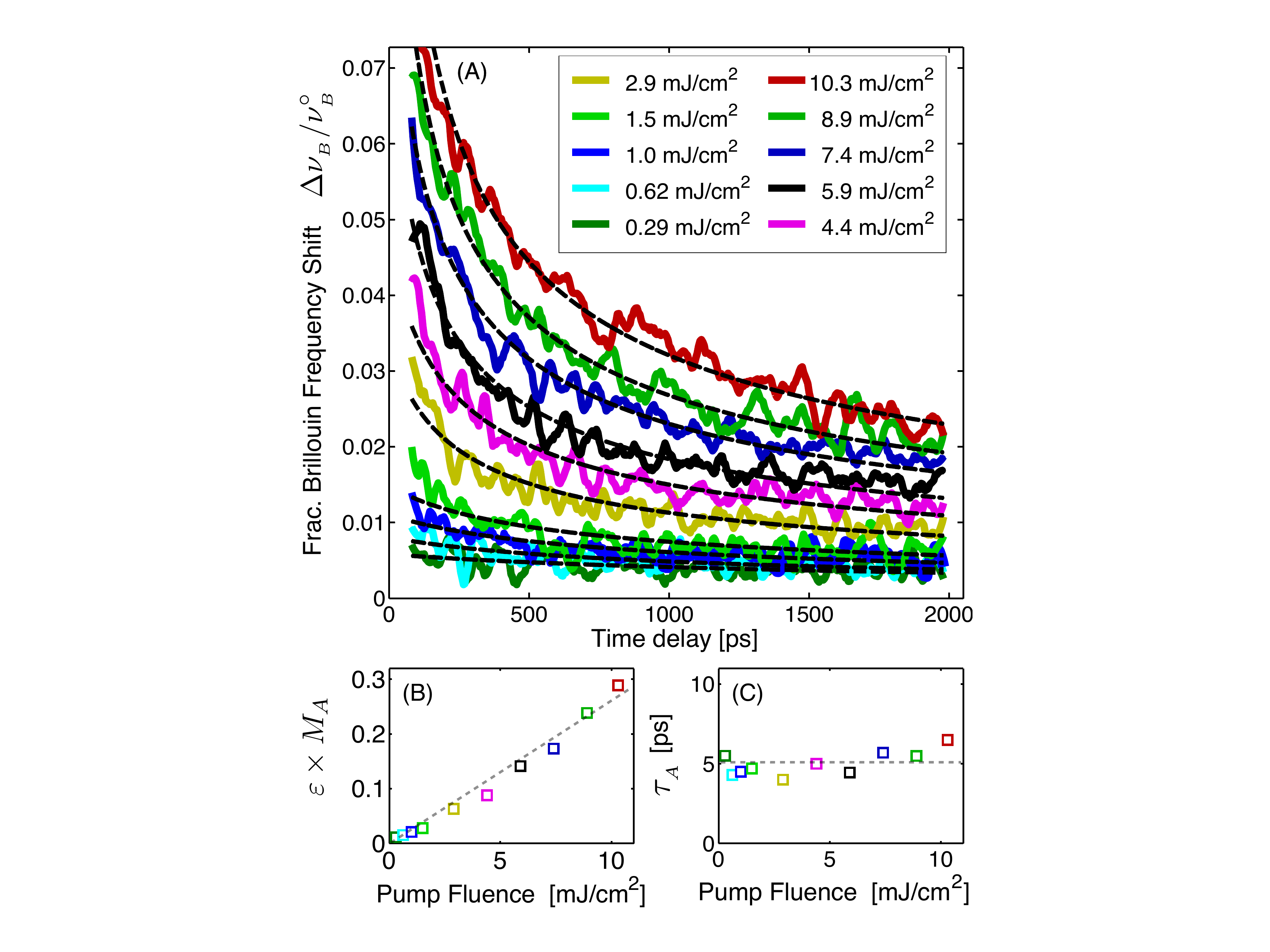}
\caption{\label{fig:data} (Color online) (A) Measured results of the fractional Brillouin frequency shift in DC704 for different laser pump fluences at 200~K sample temperature. Dashed lines are fits by Eq.~(\ref{eq:BS__freq_fitting}). (B) and (C) Fitting parameters $\varepsilon\times M_A$ and $\tau_{_A}$. Dashed lines are linear fits of the extracted parameters.}
\end{figure}

A full set of transient reflectivity data covering a broad fluence range from 0.29~$\mu$J/cm$^{2}$ to 10.3~$\mu$J/cm$^{2}$ have been acquired and analyzed. From the extracted \textquotedblleft local\textquotedblright~Brillouin frequency time variation $\nu_B(t)$, we calculated the fractional Brillouin frequency shift $\Delta \nu_{_B}(t)/\nu_{_B}^\circ$, where $\Delta \nu_{_B}(t)=\nu_{_B}-\nu_{_B}^\circ$ is the Brillouin frequency increase over the linear limit Brillouin frequency $\nu_{_B}^\circ$, see Fig.~\ref{fig:data}. We included the slight variation of the linear Brillouin frequency from a temperature change, caused by steady state heating of the sample at increasing pump fluence. From the temperature-dependent refractive index of DC704 at 395~nm, $n(T) = 1.748 - 4.9\cdot10^{-4}$~K$^{-1} \times$T~[K] \cite{KHP+13}, we have corrected the linear Brillouin frequency for the change due to laser heating according to Eq. (\ref{nu}). This resulted in a slight decrease in the linear Brillouin frequency with increasing fluence, as shown in Fig.~\ref{fig:BS_rawdata}(B). The following asymptotic equation (derived in \cite{Gusev14}), was used to fit the fractional Brillouin frequency shifts of Fig.~\ref{fig:data}(A),
\begin{eqnarray}\label{eq:BS__freq_fitting}
\frac{\Delta\nu_{_B}\left(t\right)}{\nu_{_B}^\circ} &=&\;\; \frac{\varepsilon\,M_A/2}{\sqrt{1+(\varepsilon \,M_A/2) \cdot t /\tau_{_A}}} \; ,
\end{eqnarray}
where $\varepsilon$ is the nonlinear acoustic parameter, $M_A$ the Mach number. Eq.~(\ref{eq:BS__freq_fitting}) assumes compressive strain shock pulses of triangular shape of duration $\tau_{_A}$ at FWHM and $\delta$-localized leading shock fronts, see inset Fig~\ref{fig:BS_rawdata}(A). The reduction of $\Delta\nu_B(t)$ with time in Eq.~(\ref{eq:BS__freq_fitting}) is due to the decrease of the propagating weak shock front amplitude $\sim1/\sqrt{1+(\varepsilon \,M_A/2) \cdot t /\tau_a}$ due to linear and nonlinear acoustic attenuation which tends to dissipate the nonlinear steepening of the weak shock front. The results of fitting our experimental results by Eq.~(\ref{eq:BS__freq_fitting}) with $\varepsilon M_A$ and $\tau_{_A}$ as fitting parameters are shown in Fig.~\ref{fig:data}(A). The fitting results are in excellent agreement with the datas over the complete experimental time window 100~ps~$<t<$~2000~ps, confirming the validity of our model described by Eq.~(\ref{eq:BS__freq_fitting}). From our model the emitted pulse at the aluminium transducer is rather rectangular and reshapes to a triangular shape within the first 100~ps, explaining the reason why at time delays below 100~ps the fitting is reliable but not perfect. The additional modulation of $\Delta\nu_{_B}(t)$, caused by optical interference phenomena of the probe light reflected by the trailing edge of the pulse as well as the shock front broadening can be neglected at fluences above 1.9~mJ/cm$^2$ (see Section 7 in \cite{Gusev14}). In fact our theory is better suited for higher fluences and therefore the uncertainties of the fitted parameters are negligible at high fluences. The almost linear dependence of $\varepsilon\,M_A$ on fluence, as shown in Fig.~2(B), can be used to evaluate the nonlinear acoustic parameter of the liquid at GHz frequency. Since the Mach number $M_A$ linearly scales with fluence, our results indicate that the nonlinear parameter $\varepsilon$ does not vary with fluence and can be assumed to be constant in the fluence range up to 10~mJ/cm$^2$. The mean value $\tau_A\simeq5.2$~ps in Fig.~\ref{fig:data}(C) determined from fits is in excellent agreement to the theoretically expected $\tau_{_A}\simeq$~5~ps equal to the time of sound propagation through the aluminium film, providing an additional argument in support of the model expressed by Eq.~(\ref{eq:BS__freq_fitting}). 

\begin{figure}[t!]
\centering
\includegraphics[width=8.5cm]{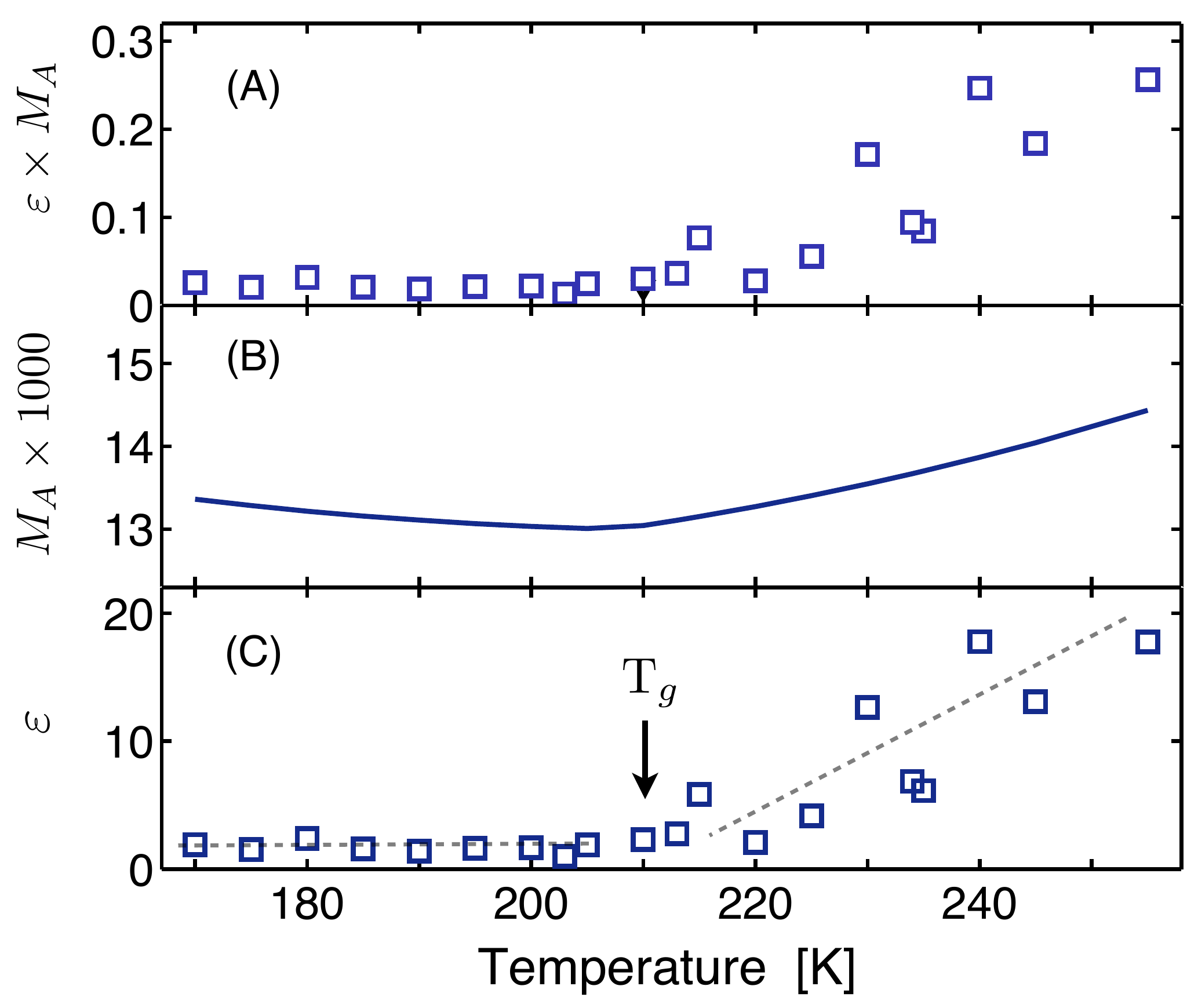}
\caption{\label{fig:temperature_dependence} (Color online) Parameter from fits to temperature dependent results at 2.9~mJ/cm$^2$ pump fluence: (A) Extracted nonlinear parameter times acoustic Mach number $\varepsilon \times M_A$, both as function of temperature, where the theoretical value of $\tau_{_A}$=~5~ps was used. (B) Calculated initial Mach number as function of temperature and (C) corresponding nonlinear parameter $\varepsilon$ versus temperature across the DC704 glass transition temperature $T_g\approx210\textrm{ K}$. Dashed lines serve as guides to the eye.}
\end{figure}

Measurements of nonlinear acoustic parameters around a glass transition have never been reported at GHz frequencies. However, we can expect that, at the glass transition, from the glass state to the highly viscous state, the acoustic nonlinearities may show a significant change due to the intrinsic structural reorganization of the glassy material. In order to check this general expectation, we have performed temperature dependent measurements of Brillouin scattering around the DC704 glass transition at 2.9~mJ/cm$^2$ pump fluence. The results of temperature variation of the fitting parameter $\varepsilon M_A$ based on Eq.~(\ref{eq:BS__freq_fitting}) are shown in Fig.~\ref{fig:temperature_dependence}. In order to extract the relevant $\varepsilon$ parameter, we have calculated the temperature dependent Mach number $M_A$ from the following equation,
\begin{eqnarray}
\label{M}
M_A = \frac{4 \ v/v_l}{(1+\frac{\rho_l v_l}{\rho \, v})(1+\frac{\rho v}{\rho_s \, v_s})} \times \beta \frac{\alpha \, F_L}{\rho \, C_p \, H} \;,
\end{eqnarray}
where $v$, $v_s$ and $v_l$ are the speed of sound of aluminium, sapphire and of DC704, and $\rho$, $\rho_s$, $\rho_l$ their densities, $C_p$ is the heat capacity of aluminium, $H$ the aluminium thickness, $\beta$ the linear thermal expansion coefficient, $\alpha$ the optical absorption coefficient for aluminum at 790~nm pump wavelength and $F_L$ the laser fluence. The second multiplier on the right-hand side of Eq.~(\ref{M}) expresses the laser generated strain while the first one is related to the acoustic transmission and reflection across the aluminium/DC704 and aluminium/sapphire interfaces. The calculation was performed with temperature-dependent values for the coefficients from \cite{Gib58,Wilson41,TAS89,KHP+13}. The calculation of the Mach number shown in Fig.~\ref{fig:temperature_dependence}(B) reveals that it is almost constant with temperature. Finally, from the calculated $M_A$, we have obtained the temperature evolution of the nonlinear coefficient $\varepsilon$ across the glass transition temperature T$_g$, see Fig.~\ref{fig:temperature_dependence}(C). As expected, the nonlinear coefficient of the glass state is lower than in the viscous state, probably because of the difference in the intermolecular interaction potential between a solid and a liquid. More surprisingly, the nonlinear coefficient increases in the liquid state about 10 times for a temperature increase of only 50 degrees above T$_g$. Our measurements highlight a significant change of the acoustic nonlinearities across T$_g$, much more pronounced than for the linear acoustic parameters, as for the speed of sound that changes only by 10$\%$ for an equivalent temperature change \cite{KHP+13}. The huge change in the nonlinear coefficient across T$_g$ has similarities with \cite{Na94} where it is reported that in ferroelectric ceramics the nonlinear parameter at 10~MHz frequency diminishes about 10 times when the temperature is diminished by 100 degrees from the Curie temperature.

Our observation of an inverse square root dependence of the shock velocity with time confirms the validity of the classical models of nonlinear acoustics \cite{RS77} extrapolated at picosecond time scales. From our measurements we have extracted material-specific nonlinear acoustic parameter which characterizes the anharmonicity of the intermolecular interaction potentials of the material under study. We determined the dependence of the nonlinear acoustic parameter of a fragile glass former DC704 at around 20~GHz at temperatures across the glass transition in the glass state and highly viscous and lightly viscous liquid state.\\


This work was partially supported by the Department of Energy Grant No. DE-FG02-00ER15087, National Science Foundation Grants No. CHE-0616939, DMR-0414895, ANR Grant Plusdil and R\'egion Pays de la Loire.

\end{document}